\documentclass[twocolumn,prb,showpacs,preprintnumbers,amsmath,amssymb]{revtex4}

\usepackage{graphicx}
\usepackage{color}
\begin{document}

\title{Specific heat 
of Ca$_{0.32}$Na$_{0.68}$Fe$_{2}$As$_{2}$ single crystals: unconventional
$s_{\pm}$ multi-band superconductivity with intermediate repulsive interband
coupling and sizable attractive intraband couplings }

\author{S.\ Johnston$^{1,2,3}$, M.\ Abdel-Hafiez$^1$, L.\ Harnagea$^1$, V.\ Grinenko$^1$, 
D.\ Bombor$^1$, Y.\ Krupskaya$^1$, C.\ Hess$^1$, S.\ Wurmehl$^1$, A.U.B.\ Wolter$^1$,  
B.\ B{\"u}chner$^{1,4}$, H.\ Rosner$^{5,1}$, and S.-L.\ Drechsler$^1$}
\affiliation{$^1$Leibniz-Institute for Solid State and Materials Research, (IFW)-Dresden, D-01171 Dresden, Germany} 
\affiliation{$^2$Department of Physics and Astronomy, University of British Columbia, Vancouver, British Columbia, Canada
V6T 1Z1} 
\affiliation{$^3$Quantum Matter Institute, University of British Columbia, 
Vancouver, British Columbia, Canada V6T 1Z4}
\affiliation{$^4$Physical Department, University of Technology Dresden, Germany}
\affiliation{$^5$Max-Planck Institute for Chemical Physics of Solids (MPI-CPfS), Dresden, Germany} 

\date{\today}

\begin{abstract} We report a low-temperature specific heat study of
high-quality single crystals of the heavily hole doped superconductor
Ca$_{0.32}$Na$_{0.68}$Fe$_2$As$_2$.  This compound exhibits bulk
superconductivity with a transition temperature $T_\textup{c} \approx 34$\,K,
which is evident from the magnetization, transport,  and specific heat
measurements. 
The zero field data manifests a significant electronic specific heat in
the normal state with a Sommerfeld coefficient $\gamma \approx 53$ mJ/mol K$^{2}$.
Using a multi-band Eliashberg analysis, we demonstrate that
the dependence of the zero field specific heat in the superconducting state 
is well
described by a three-band model with an unconventional s$_\pm$ pairing 
symmetry and gap magnitudes $\Delta_i$ of approximately  
2.35, 7.48, and -7.50 meV. 
Our analysis indicates a non-negligible attractive intraband coupling,   
which contributes significantly to the relatively high value of $T_c$. 
The Fermi surface averaged repulsive and attractive coupling strengths 
are of comparable size and  
outside the strong coupling limit frequently adopted 
for describing high-$T_c$ iron pnictide superconductors. 
We further infer a total mass renormalization of the order of five, 
including the effects of correlations and electron-boson interactions.  
\end{abstract}

\pacs{74.25.Bt, 74.25.Dw, 74.25.Jb, 65.40.Ba}

\maketitle
\section{Introduction}
Among the still  
increasing number of iron pnictide and chalcogenide
based superconductors, the hole doped systems within
the A$_{1-x}$B$_x$Fe$_2$As$_2$ (A-122) family, where A = Ba, Sr, Ca, 
and B = K, Na, and other alkaline elements, have attracted 
considerable attention. This is due to the 
availability of large stable high-quality single crystals
necessary for an accurate determination of many physical properties. 
However, within this family, the pure Ca-122 and its derived
Na-doped systems stand out in a number of 
ways when compared to  
their pure and K or Na-doped Ba-122 and Sr-122 counterparts.  

The first striking difference is the  
variation of $T_c$ with respect to doping. 
For example, 
the ``optimal" doping concentrations, where the highest values of $T_c$ 
are achieved, are rather different.  
In Ba-122 optimal doping occurs at $x\approx$ 0.3 to 0.4 
for K-doping ($T_c\approx 38.5~$K) \cite{Popovich2010} 
and $x\approx 0.4$ for Na-doping ($T_c \approx 34$~K).\cite{Aswartham2012} 
These concentrations are 
significantly smaller than the optimal $x \approx $ 0.75 ($T_c\approx 35$~K)
for Na-doped Ca-122.\cite{Haberkorn2011} 
The $T_c$ values for comparable nominal doping concentrations 
also differ significantly.  At $x \approx$ 0.5 K-doped Ba-122 
exhibits $T_c$ $\approx 36$~K vs.\ a $T_c \approx$ 18 to 19 K~for 
Na-doped Ca-122,\cite{Dong2008,Kim2012} while for the strongly doped case 
near  $x\approx 0.7$ this ratio is reversed. 
In general, the asymmetry of the electron (Co) and hole (Na) doped phase 
diagram, known also for other A-122 superconductors, is  most pronounced
for the Ca-122 family.\cite{Harnagea2011} 
La and P co-doping of Ca-122 also yields the highest
value of $T_c$ ($\approx 45$ K) among the A-122 derived superconductors.\cite{Kudo2013}  

Another difference concerns the weakly
anisotropic upper critical field.  Resistivity data  
for optimally \cite{Haberkorn2011}
and undoped (under pressure) \cite{Torikachvili2008}
single crystals of Na-doped Ca-122 
(anisotropy ratios of 1.85 $\pm$ 0.05 and 1.2, respectively), 
suggest that this system possesses the smallest out-of-plane 
anisotropy observed so far among the iron pnictides. 

The phonon spectrum in Ca-122 also displays some peculiarities. 
Anomalously large phonon linewidths have been observed by inelastic neutron scattering, 
which have been interpreted in terms of an enhanced electron-phonon (e-ph) interaction  
\cite{Phonon} relative to other pnictides. This raises the question of a 
possible role for the e-ph interaction in establishing or promoting 
superconductivity in this subgroup.\cite{Miao2013} 
These observations, combined with the presence of a 
pseudogap-like phase in under- and optimally electron 
doped single crystals,\cite{Baeck2011} and other unusual
features such as a topological Fermi surface (Lifshitz) transition,\cite{Gonelli2012}
make the Ca-122 systems a rather special 
subgroup of pnictides deserving a systematical investigation. 

From both
theoretical (electronic structure and Eliashberg-theory based analysis)
and experimental sides (ARPES) there is clear evidence that most iron
pnictide superconductors cannot be described quantitatively within the
popular two-band model and generalizations to three- or four-band models
are necessary.\cite{Popovich2010,Benfatto,Tortello2010,Ummarino} 
Another issue under debate is the total coupling strength, be it in the 
weak or intermediate vs.\ strong coupling regimes,  
and the related size of the mass renormalization in the
normal state. In the present paper we will show that 
Ca$_{0.32}$Na$_{0.68}$Fe$_2$As$_2$
with respect to these issues is also distinct from 
the nearly optimal doped Ba- and Sr-122 systems.

In this context specific heat measurements are a useful probe to 
provide key information such as the upper critical field, the
magnitude of the specific heat jump $\Delta C_p/T_c$, and the linear in 
temperature ($T$)  
specific heat (Sommerfeld) coefficient in the normal state $\gamma$$_{el}$. 
The latter reflects the strength of the electron-boson coupling 
possibly responsible for the superconductivity.
Furthermore, the low-temperature (low-$T$) specific heat can evaluate 
the presence or absence of nodes in the superconducting order parameter.  
In this paper we present low-$T$ specific heat 
measurements on single crystals of hole-doped
Ca$_{0.32}$Na$_{0.68}$Fe$_2$As$_2$.  By performing a detailed multi-band
Eliashberg analysis of the data we derive three distinct gap values for this 
system and conclude a total Fermi surface averaged coupling strength in the 
intermediate coupling regime.  
We further find evidence for a sizeable intraband component to pairing, which is
manifest as a pronounced knee in the zero field data as a function of temperature. 
Although, such investigations have been performed for
analogous compounds  
(i.e.\ K and Na-doped BaFe$_{2}$As$_{2}$),\cite{Popovich2010,Pramanik2011}  
these studies are lacking for the Na-doped CaFe$_{2}$As$_{2}$ systems.
Such systematics are needed in order to further 
clarify the differences between these structurally similar systems, 
and to determine to what extent the magnitude and symmetries of the 
superconducting order parameter, as well as the 
magnitude of the specific heat jump (coupling strength) 
at $T_c$, are sensitive to the different chemical compositions 
of the A-122 families.

\section{Methods}
\subsection{Experimental}\label{Sec:Exp}
In the present work we study thermodynamic properties of
 single crystals of the 
parent compound CaFe$_2$As$_2$ and heavily
hole-doped Ca$_{0.32}$Na$_{0.68}$Fe$_2$As$_2$. 
Single crystals of the parent system were obtained using a 
high-temperature solution-growth technique with Sn as a flux, 
similar to the one described in Ref.\ \onlinecite{Harnagea2011}.
Single crystals of Ca$_{0.32}$Na$_{0.68}$Fe$_2$As$_2$ 
were grown using NaAs as a flux.
The starting composition was selected as Ca$_{0.5}$Na$_{0.5}$Fe$_2$As$_2$:
NaAs\,=\,1:2 in a molar ratio. The mixture of the precursors CaAs, 
Fe$_2$As and NaAs were loaded in an alumina crucible. The crucible was 
sealed under argon atmosphere in a Nb container enclosed in an evacuated 
quartz ampoule. The precursor mixture was slowly heated to 1373 K, held  
there for 24 hours to ensure homogenization, and then gradually cooled down 
to 873 K at a rate of 3\,{K/h}, followed by rapid cooling to room 
temperature.

The phase purity of the resulting single crystals was investigated with 
X-ray diffraction. 
 The chemical composition was accessed 
by using a scanning electron microscope (SEM-Philips XL 30) equipped with 
an energy dispersive X-ray (EDX) spectroscopy probe. Generally, the samples 
proved to be single phase. To determine the chemical composition of Na-doped 
samples, we performed an EDX analysis for different samples 
from the same batch and at different locations on each particular sample.  
Similar to the previously reported data,\cite{Haberkorn2011} the samples 
proved to be relatively homogeneous with a standard deviation in the Na 
concentration ranging between 0.03 and 0.06. The magnetization measurements were 
performed using a superconducting quantum interference device magnetometer 
(MPMS-XL5) from Quantum Design. The temperature dependence of the electric 
resistivity of the samples was measured using a standard four-{contact} 
technique. The contacts where attached with silver epoxy such 
that the electrical current flowed parallel to the {$ab$-plane}. The heat 
capacity was measured 
with a Physical 
Property Measurement System (PPMS) from Quantum Design using a thermal 
relaxation technique down to 1.8\,K and magnetic fields up to 9\,T
applied along the crystallographic $c$-axis.

\subsection{Theory}
\subsubsection{Electronic structure calculations} 
Scalar-relativistic density functional (DFT) electronic structure calculations 
were performed using 
the full-potential FPLO code,\cite{fplo} version fplo9.01-35.
The parametrizations of Perdew-Wang \cite{PW} 
was chosen for the exchange-correlation potential within the local density (LDA). 
The calculations were carried out on a well converged mesh of 8632 k-points 
in the irreducible wedge 
of the Brillouin zone ($50\times50\times50$ mesh) to ensure a high resolution 
for details in the electronic 
density of states. The partial Ca substitution with Na was modeled within the virtual crystal 
approximation (VCA). \cite{VCA} 

\subsubsection{Multiband Eliashberg Analysis}\label{Sec:Methods_Eliashberg} 
We calculate the change in the electronic specific heat 
$\Delta C_{el}(T)$ using multiband Eliashberg theory. 
It is given by $\Delta C_{el}(T) = T\partial^2(\Delta F)/\partial T^2$, 
where $\Delta F = F_N-F_S$ is the difference between the
free energy of the system in the normal and superconducting
states.  The change in free energy can be  
expressed in terms of the mass renormalization $Z_i(\omega_n)$ and
anomalous self-energy $\phi_i(\omega_n)$ on the Matsubara
frequency axis \cite{Golubov}
\begin{eqnarray} \nonumber
\Delta F&=&-\frac{\pi}{\beta}\sum_{i,n}N_i(0)\bigg[
|\omega_n|(Z_i^N(\omega_n) - 1) \\ \nonumber
&& -2\frac{
\omega_n^2[(Z_i^S(\omega_n))^2-1]+\phi^2_i(\omega_n)}
{|\omega_n| + \sqrt{\omega_n^2(Z^S_i(\omega_n))^2 + \phi^2_i(\omega_n)}} \\\nonumber
&& +
\frac{\omega^2_nZ^S_i(\omega_n)(Z^S_i(\omega_n)-1)+\phi^2_i(\omega_n)}
{\sqrt{\omega_n^2(Z^S_i(\omega_n))^2 + \phi^2_i(\omega_n)}} \bigg]. 
\end{eqnarray}
Here, $\beta = 1/k_bT$ is the inverse temperature, 
$N_i(0)$ is the single-particle partial density of states (DOS)   
of band $i$ at the Fermi level, and the superscripts $N$ and $S$ denote 
the normal and superconducting states, respectively.
The mass renormalization and anomalous self-energy 
are obtained by solving the multi-band Eliashberg equations.
They are \cite{Benfatto}
\begin{equation}\label{Eq:Gap}
Z_i(\omega_n)\Delta_i(\omega_n) = \frac{\pi}{\beta} \sum_{m,j}
D_{ij}(\omega_n-\omega_m)\frac{\Delta_j(\omega_m)}
{\sqrt{\omega^2_n +\Delta_{j}^2(\omega_m)}}
\end{equation}
and
\begin{equation}\label{Eq:Z}
Z_i(\omega_n) = 1 + \frac{\pi}{\beta} \sum_{m,j}
D_{ij}(\omega_n-\omega_m)\frac{(\omega_m/\omega_n)Z_j(\omega_m)}
{\sqrt{\omega^2_n +\Delta_{j}^2(\omega_m)}}
\end{equation}
where $\omega_n$ and $\omega_m$ are fermion Matsubara frequencies, 
$\Delta_i(\omega_n) = \phi_i(\omega_n)/Z_i(\omega_n)$ is the gap function, and
\begin{equation}
D_{ij}(\omega_n-\omega_m) = \lambda_{ij}\int_0^\infty d\nu \frac{2\nu B_{ij}(\nu)}
{(\omega_n-\omega_m)^2 + \nu^2}.
\end{equation}
The dimensionless coupling strength $\lambda_{ij}$ 
parameterizes the coupling strength to the bosonic spectrum
$B_{ij}(\nu)$, which has both intra- [$B_{ii}(\nu)$] and interband 
[$B_{ij}(\nu)$] components. Our specific choice for the spectral densities 
and coupling constants are given in the following section. 
Finally, in order to obtain the value of the superconducting gap measured by 
spectroscopies, the self-energies are analytically continued to the 
real axis using the method of Ref. \onlinecite{MarsiglioPRB1988}. 

\section{Results}\label{Sec:Results}
\subsection{Magnetization and Resistivity}

\begin{figure}[b]
\includegraphics[width=\columnwidth,clip]{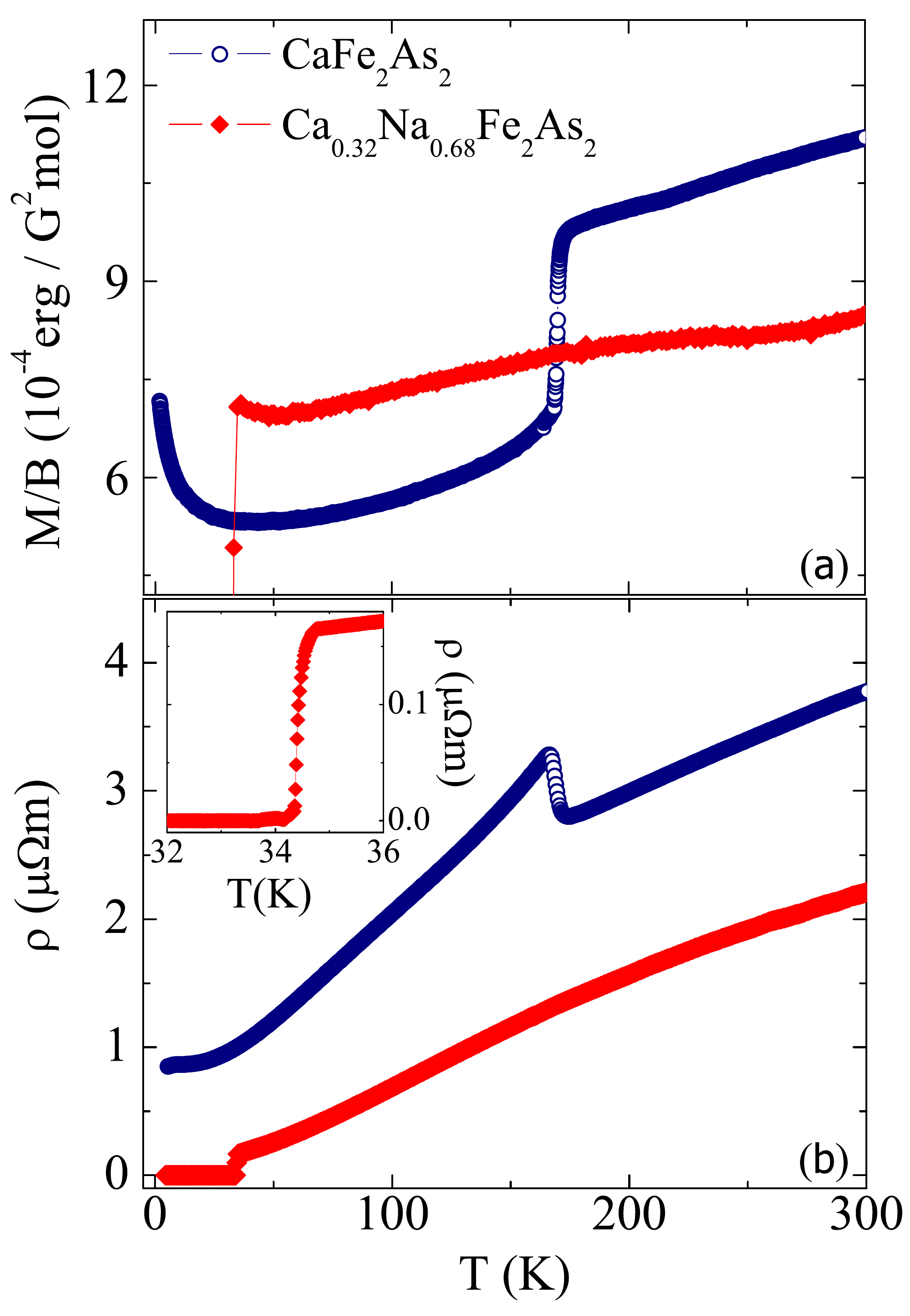}
\caption{(color online) (a) The $T$ dependence of the magnetization of CaFe$_{2}$As$_{2}$ 
and Ca$_{0.32}$Na$_{0.68}$Fe$_{2}$As$_{2}$ single crystals, measured under 
an applied magnetic field of 1~T parallel to the crystallographic basal 
plane in zero-field cooled conditions. (b) $T$-dependence of the in-plane 
electrical resistivity in zero field up to 300\,K. The inset presents 
a zoom of the resistivity data around $T_c$ for 
the Ca$_{0.32}$Na$_{0.68}$Fe$_{2}$As$_{2}$ sample.}
\label{Fig:1}
\end{figure}

Fig.\ \ref{Fig:1} shows the 
$T$-dependence of the magnetization and resistivity of 
the parent compound CaFe$_2$As$_2$ and of Ca$_{0.32}$Na$_{0.68}$Fe$_2$As$_2$ 
single crystals. Fig. \ref{Fig:1}a presents the $T$-dependence 
of the magnetization measured in zero field-cooled conditions 
and in a magnetic 
field of 1\,T applied parallel to the ab-plane. The parent compound shows a 
combined spin-density-wave (SDW) and structural transition near 169\,K, 
in good agreement with previous reports.\cite{Harnagea2011,Klingeler2010}  
The first order SDW/structural transition is completely suppressed upon 
{68\%} substitution of Ca by Na and superconductivity
appears at  
$T_\textup{c}\approx34$\,K. Fig. \ref{Fig:1}b shows the {in-plane} resistivity data 
for the samples. 
The parent compound exhibits metallic behavior over the entire 
temperature  
range with a prominent anomaly at 169\,K, in agreement with 
the magnetization data. In the Na-doped sample the SDW/structural anomaly 
is completely suppressed below $x\approx 0.5$. However,  
near $x\approx 0.7$ optimal 
doping is reached where T$_c$ is largest.
A sharp superconducting transition 
is clearly seen at $T^{\rm on}_c \approx $ 34.6\,K (90\% 
of the normal state resistivity) with 
$\Delta T_\textup{c}$ = 0.2\,K. The residual resistivity is 
$\rho(36$\,K$) \approx 17$ $\mu \Omega\cdot$cm and the residual resistivity 
ratio (RRR)
is found to be $\rho(300\,\textup{K})/\rho(36\,\textup{K}) = 12.8$. 
These values are 
similar to those reported for the other hole-doped A-122 
systems,\cite{Wang2008,Shen2011,Zhao2010} indicating a 
reasonably good quality of our single crystals.

\subsection{Specific heat}
\begin{figure}[t]
\includegraphics[width=\columnwidth,clip]{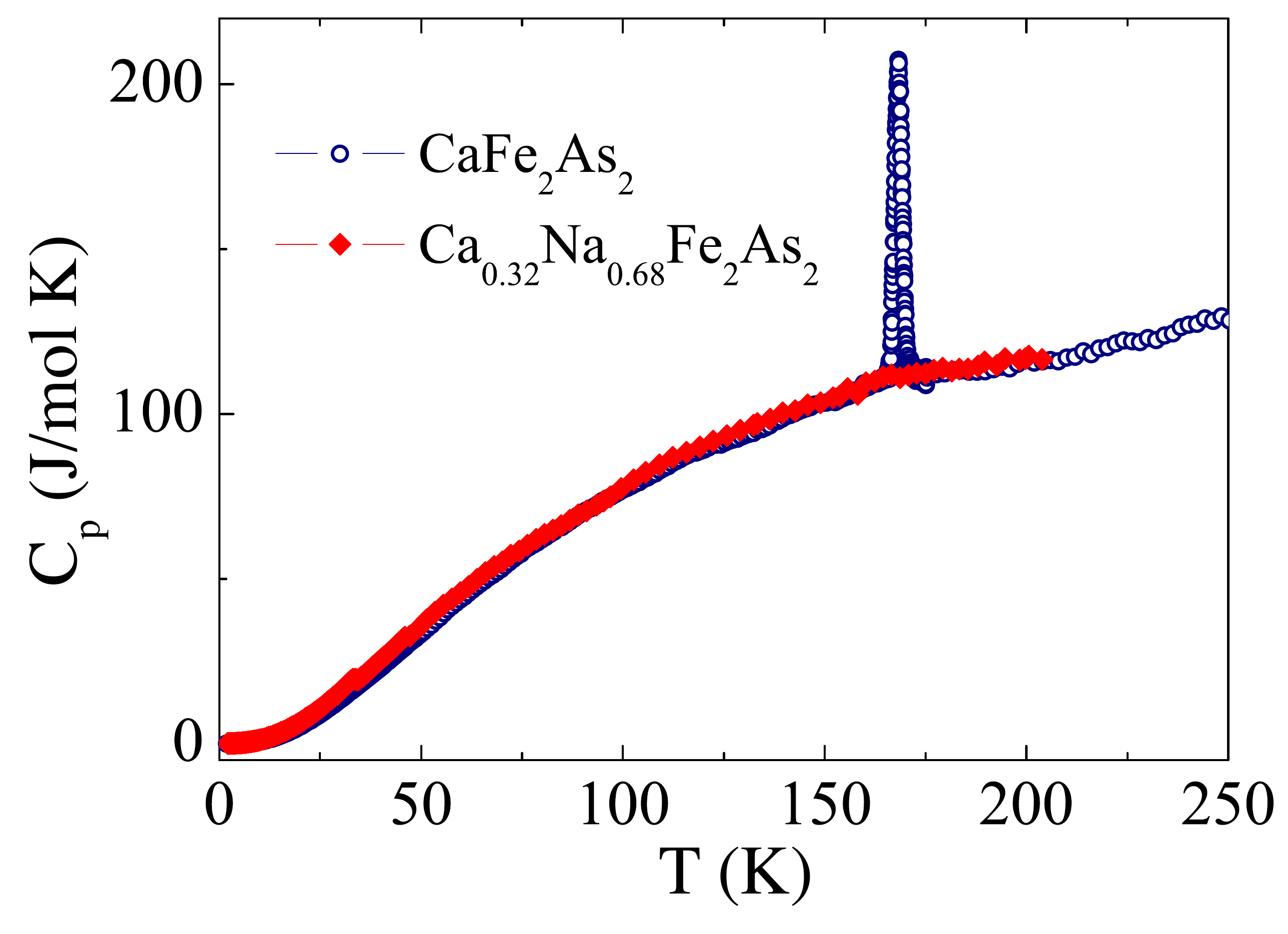}
\caption{(color online) Temperature dependence of the specific heat $C_{p}$ measured in 
zero field conditions for CaFe$_2$As$_2$ and Ca$_{0.32}$Na$_{0.68}$Fe$_2$As$_2$. 
}
\label{Fig:2}
\end{figure}

\begin{figure}[b]
\includegraphics[width=\columnwidth,clip]{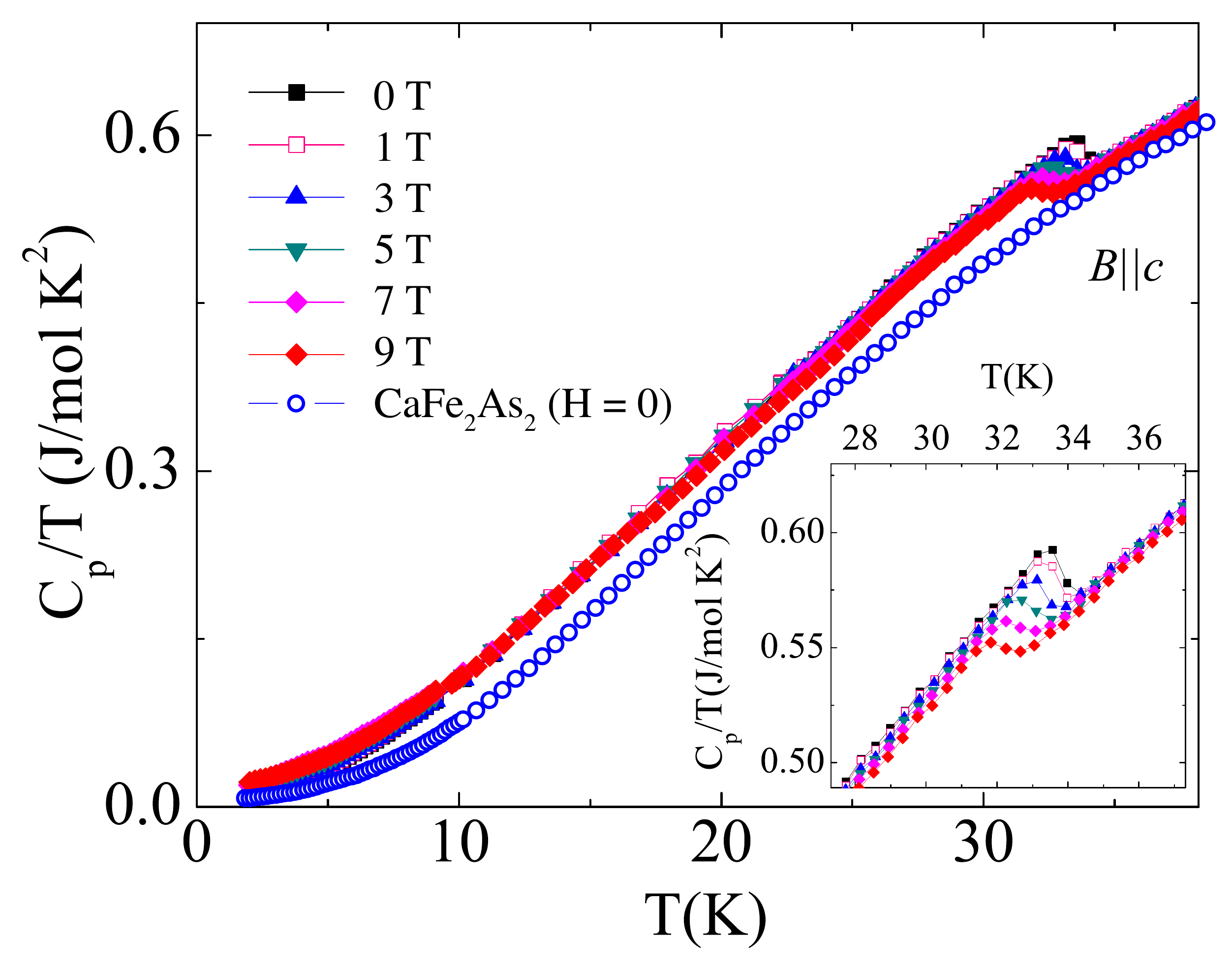}
\caption{The $T-$dependence of the specific heat 
\emph{$C_{p}/T$} of Ca$_{0.32}$Na$_{0.68}$Fe$_{2}$As$_{2}$ single crystal 
measured in magnetic fields applied along the crystallographic 
$c$-axis. For reference, the specific heat 
of CaFe$_{2}$As$_{2}$ measured in zero field conditions is also shown.  
The inset shows \emph{$C_{p}/T$} of 
Ca$_{0.32}$Na$_{0.68}$Fe$_{2}$As$_{2}$ near 
$T_{c}$. 
}\label{Fig:3}
\end{figure}

The temperature dependence of the zero-field specific heat {for} the 
parent compound and the Na-doped sample are shown in Fig. \ref{Fig:2}.  
Here again a 
sharp SDW/structural transition is observed only for 
the parent compound. The anomaly associated with the structural and 
magnetic transition is absent in the Na doped {sample}. 
Instead a jump in the specific heat associated with the superconducting 
phase transition is observed at 34\,K (see Fig. \ref{Fig:3}). 

At low $T$ the data of the parent compound 
can be fitted to $C_{p}/T = \gamma
 + \beta T^{2}$, where $\gamma$ and $\beta$ are the electronic and 
phononic coefficients of the specific heat, respectively. 
The $\gamma_{el}$ value for the parent compound 
is found to be around 5.4 mJ/mol K$^{2}$, which is in agreement with the 
 values ranging between
4.7 and 8.2 mJ/mol K$^{2}$ reported previously.\cite{Ni2008,Ronning2008} 
The phononic coefficient $\beta$ is found to be 0.508 mJ/mol K$^{4}$. 
Using the relation $\theta_D = (12\pi^{4}  R N/5 \beta)^{1/3}$, where 
$R$ is the molar gas constant and \emph{$N$} = 5 is the number of atoms per 
formula unit, we obtain a Debye temperature \emph{$\theta_{D}$} = 267\,K, 
which agrees reasonably well with previously reported data.\cite{Ronning2008} 

Fig.\
 \ref{Fig:3} summarizes the $T$-dependent specific heat measured in various 
magnetic fields for the 
Ca$_{0.32}$Na$_{0.68}$Fe$_2$As$_2$ sample, together with the 
zero-field measurements of the parent compound. The inset shows 
the specific heat data in the vicinity of the superconducting transition, 
where a pronounced jump is observed at $T_c$. 
In order to determine $T_c$ 
for each field, an entropy conserving construction has been used.\cite{Gordon1989} 


For further analysis knowledge of the electronic contribution to the 
specific heat $C_\textup{el}$ of the Ca$_{0.32}$Na$_{0.68}$Fe$_2$As$_2$ 
is required. Since this compound is non-magnetic, the total specific heat 
$C_\textup{tot}$ is a sum of the electronic $C_\textup{el}$ 
and the lattice contributions $C_\textup{ph}$. 
In the case 
of many superconductors $C_\textup{ph}$ is typically estimated by suppressing 
the superconducting transition in high magnetic fields. 
However, this option is not available here due to the high upper critical field. 
Alternatively, we have estimated $C_\textup{ph}$ using the parent compound 
CaFe$_2$As$_2$, which is not superconducting. 
As described above, the parent compound exhibits a long-range magnetic 
order of the AFM-type paired with SDW formation around 169\,K, which suggests 
a likely magnetic contribution to its specific heat. However, a recent 
inelastic neutron scattering measurement has revealed that the energy gap 
for low-energy {spin-wave} excitations in the magnetically ordered state 
is about 6.9\,meV ($\thicksim$ 80\,K) for CaFe$_2$As$_2$.\cite{McQueeney2008}  
Therefore, magnetic contributions to the specific heat should be negligible 
for $T < 40$~K {to a first approximation, }
and the total specific heat can be assumed 
to consist of only $C_\textup{el}$ and $C_\textup{ph}$ in that temperature range. 
Furthermore, for T $>$ T$_{c}$, the specific heat data of 
Ca$_{0.32}$Na$_{0.68}$Fe$_2$As$_2$ and CaFe$_2$As$_2$ samples are comparable, 
confirming similar phonon contributions to the specific heat of both samples. 
The CaFe$_2$As$_2$ {data therefore allow us to estimate} 
$C_\textup{ph}$, which can then be subtracted from   
the specific heat of our Na-doped sample. 
A similar approach was successfully applied in the case of electron 
and {hole-doped} BaFe$_2$As$_2$.\cite{Pramanik2011, Gofryk2008} 

In order to determine the phononic contribution we assume  
\begin{equation*}
 C_\textup{ph}^{\textup{CaFe}_2\textup{As}_2} = 
 C_\textup{tot}^{\textup{CaFe}_2\textup{As}_2} - C_\textup{el}^{\textup{CaFe}_2\textup{As}_2}
\end{equation*}
where $C_\textup{el}^{\textup{CaFe}_2\textup{As}_2} = \gamma_{el} T$.  
We further assume that the temperature dependence of the 
phononic contribution to the heat capacities of Ca$_{0.32}$Na$_{0.68}$Fe$_2$As$_2$
and CaFe$_2$As$_2$ are the same. Then the electronic specific heat
of the superconducting (SC) sample can be represented by 
\begin{equation}\label{eq1}
C_{el}^{SC}/T = C_{tot}^{SC}/T - \emph{f}\cdot C_{ph}^{CaFe_{2}As_{2}}/T. 
\end{equation} 
The scaling 
factor \emph{f} has been introduced to account for the difference in the 
atomic compositions {of} the parent and hole-doped compounds. The value of 
$f$ was determined from the requirement that the normal and 
superconducting state entropies at $T_\textup{c}$ be equal, i.e.,  
$\int_{0}^{T_\textup{c}}\left(C_\textup{el}/T\right)dT = \gamma_\textup{n} T_\textup{c}$, 
where $\gamma_\textup{n}$ is the normal state electronic specific heat coefficient 
for the doped superconducting sample.  
We found that the entropy conservation criterion is 
satisfied for {$f = 0.95$ (see inset in Fig. \ref{Fig:4}).

The resulting $C_\textup{el}/T$ for Ca$_{0.32}$Na$_{0.68}$Fe$_2$As$_2$  {is
presented} in the main panel of Fig. \ref{Fig:4}. The procedure yields $T_c = 33.9$\,K and 
a jump in $C_\textup{el}/T$ at $T_\textup{c}$ of $\approx$ 
39 mJ/mol K$^{2}$. 
Generally this value is higher than in the case of the electron-doped Ba-122  
compounds \cite{hardy1} and 
respectively smaller compared to the hole-doped Ba-122 
compounds.\cite{Pramanik2011, GMu2009}  
The value of the specific heat jump at $T_\textup{c}$ obtained for this 
material scales relatively well with its $T_\textup{c}$ in light of the 
recent results for the pnictide superconductors.\cite{Paglione2010} 
 
\begin{figure}[t]
\includegraphics[width=\columnwidth,clip]{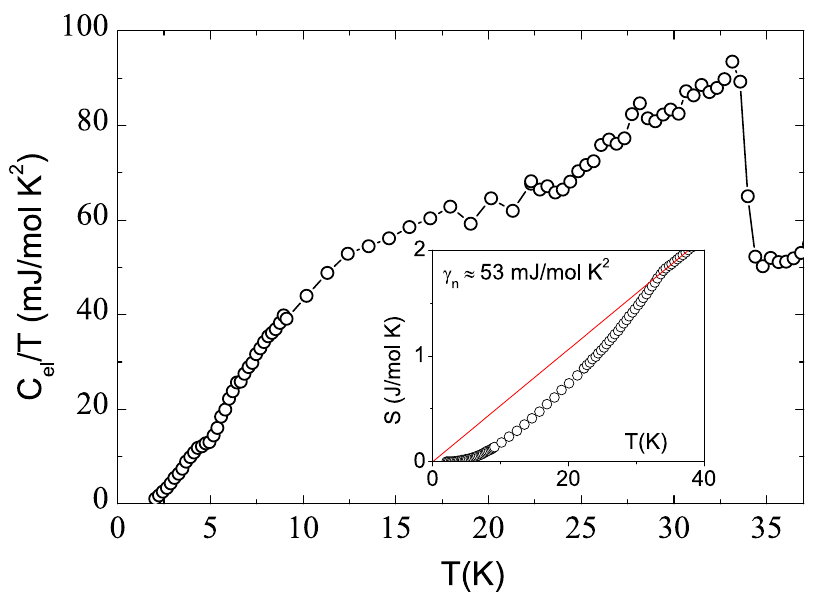}
\caption{(Color online) 
The electronic specific heat of 
Ca$_{0.32}$Na$_{0.68}$Fe$_{2}$As$_{2}$ after subtracting the phonon 
contribution as a function of reduced temperature {$t=T/T_\textup{c}$}.
In the inset the normal and superconducting state entropies are shown.
} 
\label{Fig:4}
\end{figure}
The obtained high value of $\gamma_\textup{n} = 53$ mJ/mol$\cdot$K$^2$ for 
Ca$_{0.32}$Na$_{0.68}$Fe$_2$As$_2$ as shown in Fig. \ref{Fig:4} is 
comparable to that for other members of the hole-doped A-122 series. 
From $\gamma_\textup{n} = 53$ mJ/mol K$^{2}$, we estimate 
$\Delta C_\textup{el}/ \gamma_\textup{n} T_\textup{c} = 0.96$, 
which is smaller than the prediction  of the weak coupling BCS theory 
($\Delta C_\textup{el}/ \gamma_\textup{n} T_\textup{c} = 1.43$).\cite{Poole2007}  
Taking {into} account the fact that the superconducting transition is 
relatively sharp in Ca$_{0.32}$Na$_{0.68}$Fe$_2$As$_2$, a distribution 
in $T_\textup{c}$ or the presence of impurity phases cannot explain 
the reduced value of the universal parameter (relevant
in the single-band weak-coupling case, only). In our case, however,  
this reduction is explained by the presence 
of multiple SC gaps, which can reduce the dimensionless jump
parameter in Ca$_{0.32}$Na$_{0.68}$Fe$_2$As$_2$, as evidenced in other A-122 
systems.\cite{Fukazawa2009,Pramanik2011,Gofryk2008,Hardy2010}  
Further evidence for a multi-gap scenario in this compound is 
given by the significant hump around 13\,K in our 
$C_\textup{el}/T$ vs.\,$T$ data (see
Fig. \ref{Fig:4}), which we will discuss in greater detail below.

\subsection{Multi-band Eliashberg Analysis}\label{Sec:Results_Eliashberg}
\begin{figure}[b]
 \includegraphics[width=\columnwidth]{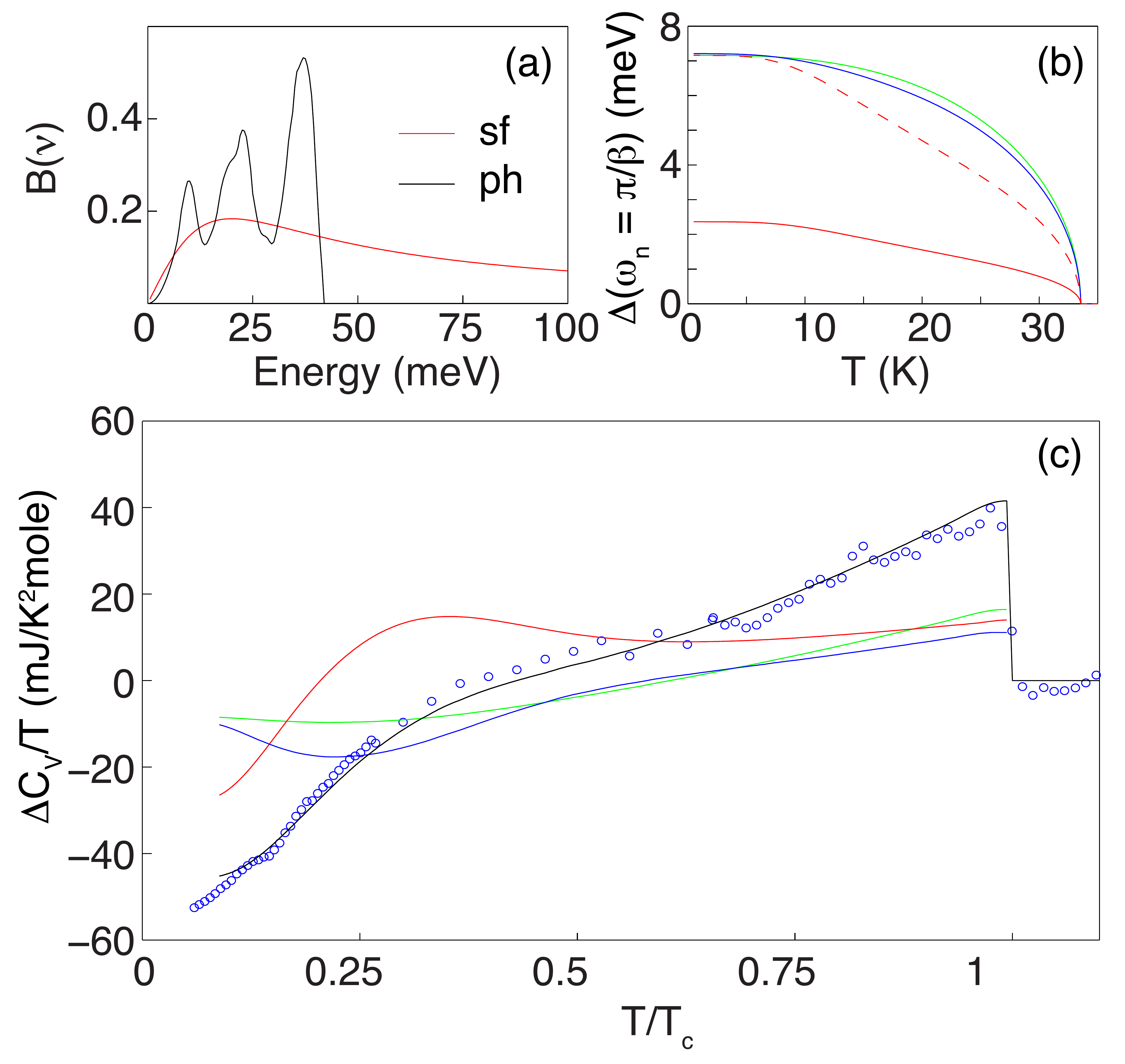}
 \caption{\label{Fig:Eliashberg} (color online) (a) The assumed
 spectrum of phonons (black/dark) and spin fluctuations (red/light).
 (b) The temperature dependence of the gap functions $\Delta_i(\omega_n = \pi/\beta)$.
 The dashed red line in (b) is a rescaled version of the solid red line, to 
 show the non-BCS-like temperature dependence of the superconducting 
 gap for this band (see main text). 
 (c) A comparison between the calculated (thick black) and measured
 (open $\bigcirc$) change in electronic specific heat $\Delta C_{el}(T)$ as 
 a function of the reduced temperature $T/T_c$.
 The individual band contributions are also shown for $T < T_c$, following
 the color scheme of panel (b). }
\end{figure}

We now undertake an analysis of the specific heat in the 
superconducting state using multi-band Eliashberg theory and 
calculate the change in electronic specific heat as outlined 
in section \ref{Sec:Methods_Eliashberg}.  
To model the self-energies we assume an effective three-band model.
One can associate two of the bands ($i = 1,2$) with different 
hole pockets and 
one band ($i=3$) with both 
electron pockets, which provide a single band by reasons of symmetry. 
(However other assignments are possible, which we discuss 
in greater detail below.) We further assume that the
intraband scattering is dominated by the attractive ($\lambda_{ii} > 0$)
e-ph interaction while the interband
scattering is dominated by a repulsive ($\lambda_{ij} < 0$)
spin fluctuation mediated interaction.\cite{Popovich2010} 
(Note that the negative sign for $\lambda_{ij}$ only
enters into Eq. (\ref{Eq:Gap}) while in Eq. (\ref{Eq:Z}) all $\lambda_{ij}$
enter with a $+$ve sign.)
The bosonic spectral densities $B_{ii}(\nu) = B_{ph}(\nu)$ and
$B_{ij}(\nu) = B_{sf}(\nu)$, respectively, are shown in
Fig. \ref{Fig:Eliashberg}a. The phonon spectrum is taken from
Ref.\ \onlinecite{Phonon} while the spin fluctuation
spectrum is assumed to have the form
$B_{sf}(\nu) = \Omega_{sf}\nu/(\nu^2+\Omega_{sf}^2)$ ($\Omega_{sf} = 20$ 
meV).\cite{Benfatto,Millis}  In both cases $B_{ij}(\nu)$ has been normalized
such that $\int_0^{\omega_c} d\nu 2 B_{ij}(\nu)/\nu = 1$ and the spin
fluctuation spectrum has been cut off for $\nu > \omega_c = 100$ meV. 
We solve Eqs. (\ref{Eq:Gap}) and (\ref{Eq:Z}) self-consistently assuming
an $s_{\pm} $ gap symmetry and treating the values of $\lambda_{ij}$ and 
$N_i(0)$ as adjustable parameters. 

The $T$-dependence of the gaps $\Delta_i$ at the $n=1$ Matsubara frequency are 
shown in Fig. \ref{Fig:Eliashberg}b.  Our model gives  
$T^{\rm fit}_c$ = 33.6 K and the low temperature ($T = 0.5$ K)
gaps are $\Delta_i(\omega_n = \pi/\beta) = 7.16$, $2.36$, and $-7.20$ meV.
{The corresponding spectroscopic gaps on the real axis
are $\Delta_i= 7.48$~meV, 2.35~meV, and -7.5~meV, respectively.
The non-BCS temperature dependence for the smallest gap  $\Delta_2$ 
is a typical feature of the outer FSS ($h_3$) which are very weakly coupled to 
the remaining ``strongly" coupled  
FSSs $h_1$, $h_2$ and $e_1$, $e_2$. Thus, its observation points to the need 
for a multiband (three or more) model in order to obtain a correct 
assignment of the bands and their respective couplings.} 
 
The total change in the electronic 
specific heat is compared to the experimentally determined data (open circles)
in Fig. \ref{Fig:Eliashberg}c. 
The agreement between the Eliashberg model and the data is good given the
simplicity of the model. However, we obtain a total electron-boson coupling in
the intermediate regime. 
{The fitted values of the partial DOS for each band
are (in eV$^{-1}$) $N_1(0) = 0.71$,
$N_2(0) = 3.80$, and $N_3(0) = 0.59$.}  
The dimensionless coupling constants are $\lambda_{11} = \lambda_{22} =
\lambda_{33} = 0.45$, 
$\lambda_{23} =-0.1$, $\lambda_{13} =-1.0$, $\lambda_{12}=0$, 
and the interband balance relation 
$\lambda_{ji}N_j(0) = \lambda_{ij}N_{i}(0)$ was imposed. 
These values result in a total average coupling  
$\lambda_{av} = \sum_{ij} (N_i(0)/N_{tot})|\lambda_{ij}| = 
\lambda_{ph}+2\left(|\nu_1\lambda_{13}+\nu_2\lambda_{23}|\right) = 0.88$, 
where the {\it phenomenological} normalized 
partial DOS   
$\nu_\alpha=N_\alpha(0)/N_{tot}(0)$ have been introduced. 
This value is significantly {\it smaller} than 
the average coupling constant $\lambda_{\rm av} \approx 2$ obtained  
for Ba$_{0.68}$K$_{0.32}$Fe$_2$As$_2$ 
and other pnictides.\cite{Popovich2010}  
Finally, if we neglect the $e$-$ph$ interaction we obtain $T_c = 21.7$ K. 
Similarly, if we neglect the spin fluctuations we obtain a $T_c = 5.4$ K. 
This indicates that the phonons produce a non-linear enhancement of $T_c$ when 
operating in conjunction with a dominant spin fluctuation  
mechanism, similar to that proposed for the cuprates.\cite{SJohnston2010}

\begin{figure}[b]
\includegraphics[width=12pc,clip]{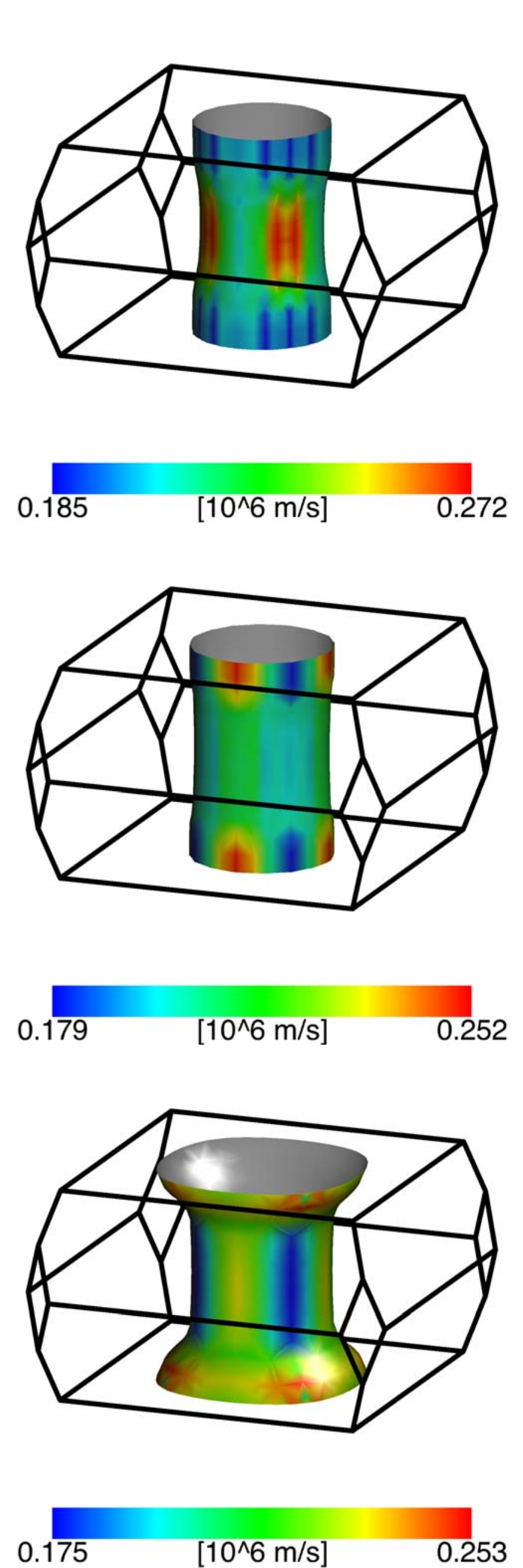}
\caption{(color online) The three hole Fermi surface sheets (FSSs) around the 
$\Gamma$-point formed by the bands $h_1$ (upper), $h_2$ (middle), and 
$h_3$ (bottom) according to our LDA calculation.
The color denotes the magnitude of the Fermi velocities.
Within scenario I we tentatively assign the upper ($h_1$) 
and the middle ($h_2$) FSSs 
to band ``1"
in our three-band Eliashberg analysis (see subsection D) 
and the lower FSS ($h_3$) to band 2.
Within scenario II the 
band $h_2$, only, forms band ``1",
whereas the two remaining 
hole bands $h_1$ and $h_3$ to
form the effective band ``2".  
The color scale denotes the magnitude of the
calculated Fermi velocities.
\label{Fig:HoleFS}}
\end{figure}

{As mentioned previously, multiple scenarios can 
be considered when making assignments between our effective threeband model 
and the five bands crossing the Fermi level in the real material. In the following 
section we will discuss these scenarios in the context of our our 
DFT calculations. }

Scenario I:  
One naively takes the $i = 1$ band to represent a combined contribution from 
the inner two hole bands (denoted $h_1$ and $h_2$), while the $i = 2$ band 
represents the third hole band $h_3$ and 
$i = 3$ represents the combined electron pockets $e_{1,2}$.  

Scenario II: 
The second hole band $h_2$ and the outer electron
band $e_1$ form the strongly coupled bands ``1" and ``3" that 
are responsible for the $s_{\pm}$ symmetry, while the remaining two hole
bands $h_1$ and $h_3$, together with the second electron band $e_2$, form 
the effective weakly coupled band ``2".

Scenario III: The two electron bands
$e_1$ and $e_2$
would belong to different symmetries or their gap structure is 
highly anisotropic with a large value on the outer part
$e_1$ and the small value on $e_2$. 
Then electron band with the small gap, 
together with the hole bands $h_1$ and $h_3$, would lumped in that effective weakly coupled
band ``2". Alternatively, the large PDOS of 
band 2 should be ascribed to a specific high-energy renormalization
acting on the orbitals contributing essentially to the largest hole 
band $h_3$.

{Recent ARPES measurements\cite{Evtushinsky2013} have observed 
an additional hole-type FSS near the X-point sometimes 
denoted as the ``propeller blade" or $\varepsilon$-FSS.  
We would like to suggest that this Fermi surface may be 
affected by surface effects. Such a point of view
is supported by a significant difference in the cross sections 
reported
for an analogous propeller FSS in KFe$_2$As$_2$, as observed by 
ARPES \cite{Yoshida2012} and dHvA \cite{Terashima2013} measurements  
where a ratio $\sim 2$  (i.e.\ from 1.91 to 2.44) has been reported. 
Since the dHvA data  
expected to measure bulk properties, one might conclude that the ARPES
data exhibit a larger propeller cross section due to 
a relaxed lattice structure near the surface. Within LDA
the propeller Fermi surface occurs in the very vicinity of 
stoichiometry KFe$_2$As$_2$, and the corresponding  
Lifshitz transition for the $\varepsilon$-FSS might take place at higher
doping ratios beyond the optimal doping at around 0.7. In this context
the lower $T_c$ of about 31~K for the sample investigated in Ref.\ 
\onlinecite{Evtushinsky2013} might be relevant. Further theoretical and 
 experimental studies are necessary to settle this issue, which is 
of interest for all hole doped 122 systems.}

\subsection{Theoretical Aspects of Normal State Properties}
We now consider the intermediate coupling strength inferred from our 
Eliashberg analysis in the context of the total mass enhancement 
for the Ca-122 system.  For this purpose, the DOS  
$N(\omega)$ and the Fermi surface (FS) topology are required.  
We will adopt the electronic structure 
calculated within density functional theory and the 
local density approximation (LDA) as 
close to a bare electronic structure without important many-body 
corrections (see Fig. \ref{DOS}).
Our aim is to compare the empirically determined Sommerfeld 
constant $\gamma_{el}$ in the electronic specific heat to 
the value bare value $\gamma_b$ determined from the bare 
reference DOS at the Fermi level N($\omega=0$) 
\begin{equation}
\gamma_b=\frac{\pi^2k^2_{\mbox{\tiny B}}}{3}N(0)=2.5352N(0),
\end{equation}
where $\gamma_b$ is measured in units of mJ/f.u.$\cdot$K$^2$ 
and $N(0)$ is measured in states/eV$\cdot$mol. In this way we can 
estimate the total mass enhancement due to high-energy 
correlations and electron-boson coupling. 

\begin{figure}[]
\includegraphics[width=\columnwidth]{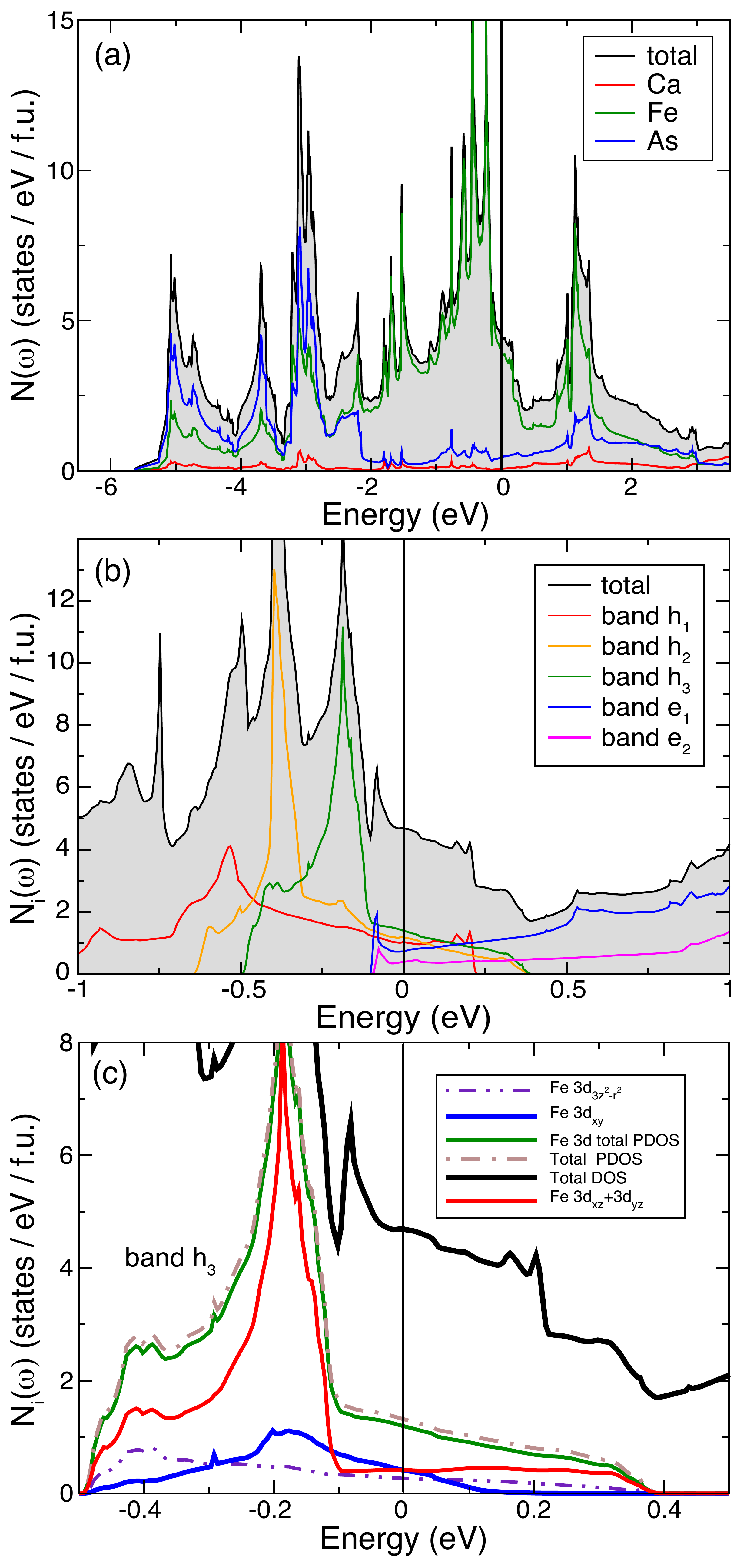}
\caption{(color online) (a) 
The electronic density of state (DOS) from LDA-FPLO calculation for
Ca$_{0.32}$Na$_{0.68}$Fe$_{2}$As$_{2}$ with the various 
elemental contributions from Fe 3$d$, As 4$p$, and Ca 4$s$ states.
(b) The Fermi surface sheets (FSSs) band-resolved contributions.
Bands $h_{1-3}$ include the three hole-type FSSs centered 
around the $\Gamma$-point
(see Fig. \ref{Fig:HoleFS}) 
and $e_{1,2}$ are the two electron-type FSSs (see Fig \ref{Fig:ElectronFS}), 
respectively.
(c) the orbital resolved partial DOS for the outer hole 
FSSs $h_3$ (see Fig.\ \ref{Fig:HoleFS}), lower panel).
The 3$d_{xz}$ and 3$d_{yz}$ are degenerate in the present tetragonal symmetry and 
have the same weight. The notation of orbitals is the same as that used 
in the ARPES
and dHvA -literature for Fe pnictides, i.e.\ there is a 45$^{\circ}$ rotation
of the $x$ and $y$ axes with respect of the original tetragonal axis of the 
 the Bravais cell.   }
\label{DOS}
\end{figure}

\begin{figure}[b]
\includegraphics[width=13pc,clip]{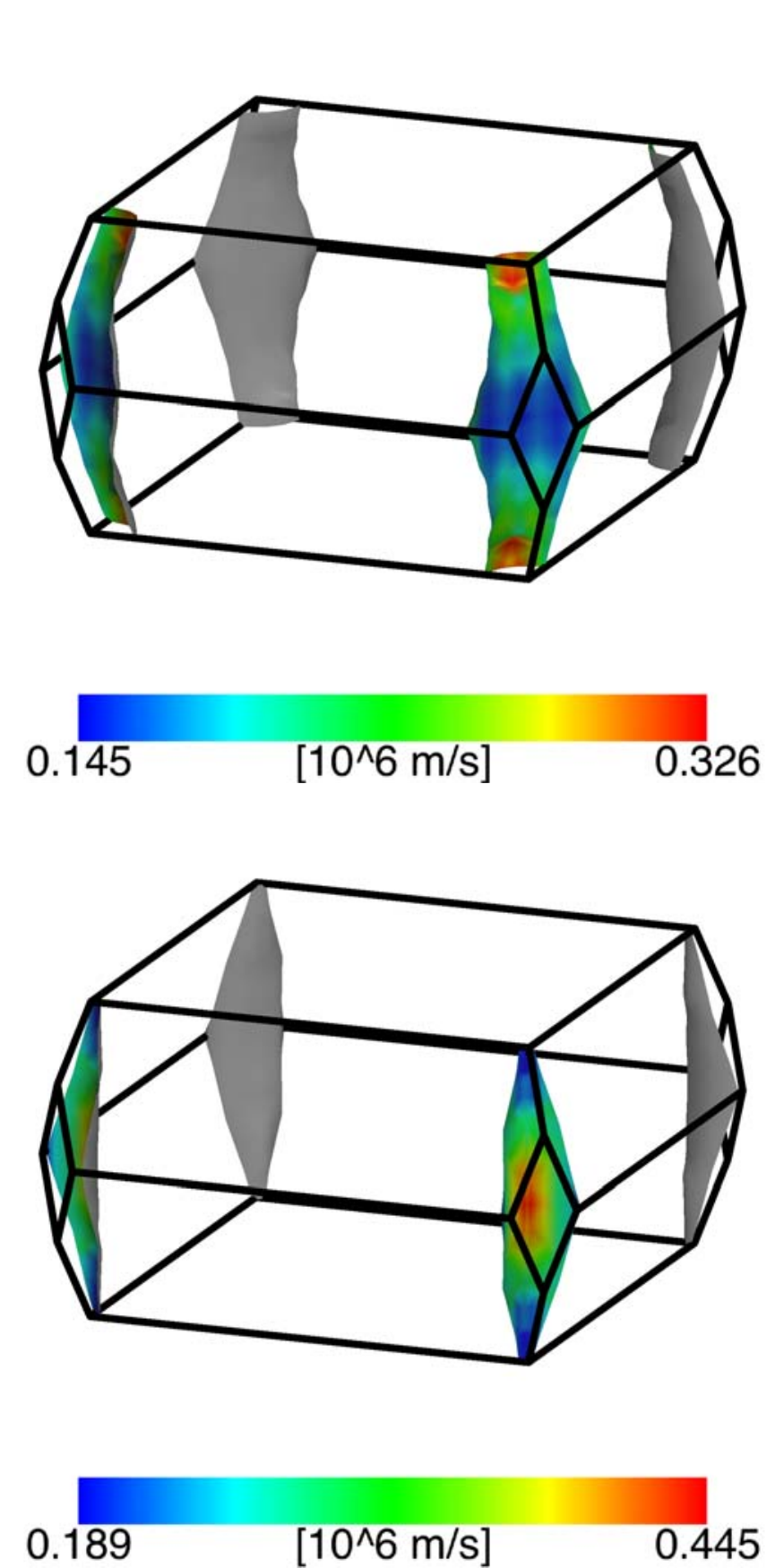}
\caption{(color online) The two electron FSSs resulting from bands $e_1$ and $e_2$
according to both scenarios I and II 
are tentatively assigned to 
to the effective band 3 in our Eliashberg analysis of the specific heat 
(see subsection D).}
\label{Fig:ElectronFS}
\end{figure}
Using the structural data for Ca$_{0.34}$Na$_{0.66}$Fe$_2$As$_2$,\cite{ZhaoK2011} 
being very close to the composition of our single crystal, we arrive at 
$N(0)=4.69$ states / eV / f.u. or $\gamma_b =10.96$~mJ/K$^2$mol. 
Here, the Na doping has been treated within the standard virtual crystal approximation. 
From our calculated 
plasma frequencies of 3.15~eV (in-plane) and 1.94~eV (out-of-plane), 
a relatively small mass anisotropy of 2.64 is found.  From this we 
estimate an anisotropy for the upper critical fields of 1.63 
(in a single-band approximation), which compares reasonably well with 
available experimental values of 1.85$\pm 0.05$ for $x=0.75$
and 1.82 for $x=0.5$, respectively.\cite{Haberkorn2011}  
The total mass enhancement factor can estimated from the specific heat data 
reported here  
as $\gamma_{el}/\gamma_b \approx 4.95$. This value compares very well
with that obtained from the experimental (significantly renormalized) Fermi velocities of 
5.06 $\times 10^6$ cm/s for both $x=0.5$ and $x=0.75$ \cite{Haberkorn2011,footnote3} 
when 
compared to the averaged bare Fermi velocity of 2.44 $\times 10^7$ cm/s obtained from 
our LDA calculations. 

The mass enhancement can also be estimate from the slope of the upper critical field. 
From our data we find 
$ -dH_\textup{c2}^\textup{(c)}/dT = 5.5$\,T/K which exceeds the reported
values of 4~T/K \cite{Haberkorn2011} and 4.5~T/K \cite{ZhaoK2011}
for $x=0.75$ and $x=0.66$, respectively (obtained from 
resistivity measurements). 
For our value of -5.5~T/K for $H^c_{c2}$,
a renormalization factor of 5.56 for $x=0.68$ is obtained, which is comparable to 
the values 4.83 derived from the data of Ref.\ \onlinecite{Haberkorn2011} 
and 5.02 derived from Ref.\ \onlinecite{ZhaoK2011}. Thus we estimate the 
total mass renormalization for Na-doped Ca-122 to be of order five.  

The total mass renormalization is the product of both the low-energy 
electron-boson (e-b) and high-energy electron-electron (e-e) 
interactions. In order to isolate approximately the contributions from the two 
we employ a simple factorization ansatz \cite{Footnote2} 
\begin{equation}
\gamma_{el}=\gamma_b \left( 1+\lambda_{\rm b, tot} \right)
 \eta ,
\label{product}
\end{equation}
where $\eta=1+\lambda_{\rm e-e}$ describes the effective high-energy 
renormalization from Coulomb and Hund's interactions that are responsible for the
well-known band narrowing observed in photoemission spectroscopies. 
The low-energy effects from the interaction of conduction 
electrons with bosonic excitations, such as low-energy spin-fluctuations or  
phonons, contribute to $\lambda_{\rm b, tot}$. 
If we adopt a usual high-energy 
renormalization for 3$d$-transition metal compounds 
$\eta \approx 2.5$ - $3$,\cite{Qazilbash2009,Yin2011}  
say 2.6 - 2.7\cite{footnote1} 
then we are left with a constraint for the total bosonic
energy renormalization ranging from 1.738 to 1.902,
 i.e.\
\begin{equation}
\lambda_{b,\rm tot} \approx 0.82 \pm 0.15
\quad ,
\end{equation}
for the total electron-boson coupling constant
averaged over all Fermi surface sheets (FSS). 
Such a value of ${\lambda}_{\rm b,tot}$ is in the 
intermediate coupling regime and agrees well with the 
value obtained in our multi-band Eliashberg analysis. 
The high-energy renormalization $\eta$, which stems from the 
sizeable Coulomb interaction
and/or the Hund's rule coupling (see e.g.\ Ref.\ \onlinecite{Yin2011}), 
is outside the energy region treated in the standard Migdal-Eliashberg theory. 

We note that since the title compound is considered within a multi-band
approach, adopting a single parameter for the 
description of the high-energy renormalization means that all bands are 
considered to be renormalized uniformly in the same way. However, in case of 
correlated multi-orbital systems like the A-122 systems this can 
be violated. For example,  
bands with a significant contributions from the 3$d_{xy}$ orbitals 
can show a much stronger renormalization close to a Mott transition 
(orbital selective Mottness).\cite{Medici2013}

\section{Discussion}
A number of additional comments regarding our Eliashberg result are in order. 
First, the partial DOS for the second
hole pocket, which we associate with band $h_3$ in scenario I 
is rather large: it is twice as large  as the relative weight
from our LDA results (see Figs.\  \ref{Fig:HoleFS} and \ref{Fig:ElectronFS}). 
The discrepancy can be considerably reduced if the $h_3$ band 
would be upshifted by $\approx 100$~meV. Since the $h_3$ band is dominated by 
$3d_{xy}$-states, it is most sensitive to many-body effects beyond the LDA.  
If one assumes a twice as large  
partial DOS for the $h_3$ band, the total bare DOS would increase too, 
reducing our estimate for the 
the total mass renormalization to 3.71 (see Eq.\ 2). Then, adopting a
slightly smaller high energy renormalization $\eta =2$, we again arrive at 
a bosonic factor of 1.85, i.e.\ a total coupling constant $\lambda_{\rm tot}=0.85$
in accord with our Eliashberg-theory result of 0.88.
Optical measurements yielding the renormalized plasma frequency would
be helpful to support such a scenario.
Comparing the empirically obtained partial DOS 
obtained from our model to those obtained in our LDA 
calculations 
we again estimate the high-energy $\lambda_{\rm e-e} \approx 1.2$, 
i.e.\ somewhat smaller than 1.6 suggested from the adopted values 
for the plasma frequency. 
From the other side a ``corrected" band structure, with slightly shifted
bands introduced to reproduce the FSS cross sections, might result in somewhat 
different numbers. Adopting now a possibly more realistic ``average" value
of  $\lambda_{\rm e-e} \approx 1.4$, one would arrive at 
$\lambda_{\rm e-b} \approx 1.06$, which is of the same order as 
the value obtained above and still well within the intermediate coupling regime.

Second, the relatively weak coupling between the second effective hole pocket
and the electron bands, as well as the relatively large contribution from the
intraband coupling, implies that the weakly coupled band becomes 
superconducting at $T = T_{\rm c}$ via a proximity effect.  In the absence of
interband coupling this band would have a much lower $T_{\rm c}$.  This
situation was required in order to reproduce the pronounced knee in $\Delta C_{\rm
el}$ near $T/T_{\rm c }\approx 0.3$.  If a strong interband coupling is assumed
between these bands, or if the intraband coupling is reduced, this knee 
is significantly muted.\cite{Popovich2010,Benfatto} 
 
Third, the averaged 
intraband coupling constant of $\lambda_{\rm ph} = 0.45$ is comparable to the 
repulsive interband counter part of $\lambda_{\rm sf} = 0.446$. The former exceeds 
the estimates of $\lambda_{\rm ph} \approx 0.2$  based on standard LDA-based calculations
for the La-1111 system \cite{Boeri2008} by slightly more than a factor of two. 
At the moment it is unclear to what extent this is a specific property of 
the Ca-122 compounds or a more general many-body driven
enhancement e.g.\ {vertex corrections for the e-ph interaction 
due to orbital fluctuations 
\cite{Kontani2010} or} residual correlation effects beyond the LDA, 
as suggested for the high-$T_c$ 
cuprates.\cite{Roesch2004,Mischenko2004,SJohnston2010,Johnston2012} 
In the context of the Ca-122 derived systems it is also noteworthy that a 
somewhat enlarged electron-phonon
coupling constant $\lambda_{\rm ph} \approx 0.37 - 0.38 $
has been reported
for Co-doped Ca-122 from first principle calculations.\cite{Miao2013}Finally, 
a different symmetry of the order parameter (beyond the scope 
of the present paper) such as $s_{\pm} +id$ or nodal $d$-wave 
symmetry are expected at least for the vicinity to
the overdoped case near $x=1$, where all electron Fermi surfaces might disappear. 
Such a case probably requires a somewhat larger $\lambda_{\rm av}$ to reproduce the 
same experimental $T_c$.  


\vspace{0.5cm}

\begin{table}[b]
\caption{
Empirical and LDA partial density of states $\nu_i(0)
=N_i(0)/N(0)\ i=1,2,3 $ at the Fermi level 
(PDOS) for {scenarios I, II, and III} discussed in
the text.}
\vspace
{0.3cm}
\begin{tabular}{|l|l|l|l|l|} \hline
PDOS
 & Eliashberg fit& I&  II&  III 
\\ \hline\hline
$\nu_1(0)
$& 0.139 &0.467&0.250& 0.250 \\
$ \nu_2(0)
 $&0.745 & 0.296 & 0.513 & 0.595 \\
$\nu_3(0)
$ &0.116 & 0.237 & 0.237 &
0.155\\
$\nu_1(0)/ \nu_3(0)
$ & 1.203 &1.974 & 1.057 & 1.404 \\ \hline
\end{tabular}
\end{table}

\vspace{0.5cm}

In the context of a non-negligible intraband electron-phonon
interaction the recent four-band analysis of LiFeAs by Ummarino 
{\it et al.} \cite{Ummarino} is very interesting. These authors 
calculated first the phonon 
part of the Eliashberg function and treated the e-ph
coupling strength as a fitting 
parameter, similar to our approach. But the resulting significant intraband 
interaction was assigned to a single band, only, 
in contradiction with recent ARPES data, 
\cite{Borisenko,Kordyuk} which points to phonon features on all four FSSs. 
Furthermore, the partial
DOS $\nu_i=N_i(0)/N(0)$
were adopted from the LDA-calculations, 
at variance with our approach treating them as adjustable parameters. 
In order
to reproduce the observed largest gap of 5~meV on the inner hole-pocket's 
FSS, a relatively large intraband coupling constant $\lambda_{11} \approx 0.9$
was required. The authors interpreted their results as a 
fictitious effect due to the violation of Migdal's theorem in narrow bands 
with small Fermi energies. 
{Here, recent NMR measurements in the oxygen free Co-doped
Ca-1111 systems are of interest, since 
the presence of both spin and orbital fluctuations 
has been claimed in the interpretation of the data.\cite{Tsutsumi2012}}
The relevance of a sizeable intraband interaction has been 
stressed in Ref.\ \onlinecite{Kulic2009}. Finally, the recently discussed
non-magnetic impurity driven $s_{\pm} \rightarrow  s_{++}$ transition at 
a still sizeable $T_c$ \cite{Efremov2011} makes sense for a considerable 
intraband coupling.

\section{Summary and Conclusions}
We have examined the electronic specific heat
data of good-quality single crystals 
of Ca$_{0.32}$Na$_{0.68}$Fe$_2$As$_2$ ($T_\textup{c} = 34$\,K).  
The low-temperature data in the superconducting state is  
well described by an effective three-band model with {an 
$s_{\pm}$-symmetry for} the superconducting order parameter 
and comparable Fermi surface averaged intra- and interband coupling 
strengths. {From our model we obtain} gap 
values of $|\Delta| \approx $ 2.35, 7.48, and 7.5~meV{. 
This is in close agreement with   
recent ARPES measurements \cite{Evtushinsky2013} where 
2.3 and 7.8 ~meV were reported for the outer 
and the inner FSSs, respectively.}
However, the same large gap has been observed in ARPES
measurements for an addtional hole-like FSS, which we ascribe to a surface
induced feature beyond the scope of the present paper.  
{However, it should be noted that the 
magnitude of the gaps on the inner hole and inner electron
FSSs are difficult to resolve experimentally within ARPES. 
Within the different scenarios for the band assignments considered here, 
different interpretations
are possible. In scenario I, large gaps would be present on the inner 
hole FSSs $h_1$ and $h_2$ and on the two electron FSSs $e_{1,2}$ while 
the small gap would appear on the outer hole FSS $h_3$. 
In scenario II the large gaps would appear on $h_2$ and $e_2$ while the smaller 
gap appears on the remaining bands. A similar assignment would be made for scenario III. 
Scenario II provides the best agreement with the LDA-derived ratios for 
the PDOS (see Tab.\ 1). 
However, the presence of a symmetry breaking mechanism making
$e_1$ and $e_2$ nonequivalent remains unclear at present.
Other experimental probes such as optical conductivity might be helpful to resolve this
question, along with further theoretical calculations within a four or five-band models.}

The remaining small deviations of about 0.3~meV between   
the two large gap-values reported in Ref.\ \onlinecite{Evtushinsky2013}
and our predictions might be ascribed to a slight 
enhancement of the interband coupling at low-temperature as 
expected in a self-consistent
$T$-dependent treatment of the spin fluctuations (i.e.\ the 
formation of a resonance mode in the superconducting state)   
and/or some gap anisotropy. Observations of the latter have been made 
for the outer hole-type FSS. Both effects are neglected in our current 
approach.

From our fit of the electronic specific heat we obtained a  
total electron-boson coupling constant, averaged over all Fermi 
surface sheets, with $\lambda_{\rm tot} \approx 0.9$. 
Furthermore, a sizeable amount of the total coupling 
is provided by intraband e-ph coupling with 
$\lambda_{\rm ph}\approx 0.45$. This value is enhanced compared to the value 
typically obtained by density functional theory calculations. This implies that the 
value of $T_c$ in the Na-doped Ca-122 systems is enhanced by 
the attractive intraband coupling.  
The value of $\lambda_{tot}$ is in excellent agreement with the value estimated 
from the renormalized Sommerfeld constant of the electronic 
specific heat of about 53~mJ/K$^2$mol (f.u.). This points to moderate or
weak coupling with $\lambda \approx 0.9 $, only, as discussed in section 
III.C. We stress once more that the high-energy renormalization  
typical for itinerant 3$d$ metals yields a larger contribution
($\approx 2.6$) than the total bosonic one ($1 + \lambda_{\rm b} \approx 1.82\pm 0.15$, 
or $\lambda_{\rm b} < 1$). 

Finally, both the specific heat and the upper critical 
field data provide
a significant total mass renormalization of the order of five including
both bosonic and high-energy renormalizations.  
For a full understanding of the gap structure and the nature of 
superconductivity of the Na-doped Ca-122 system, further studies on 
materials with different doping levels are required.\\
 
\vspace{0.5cm}
\begin{acknowledgments}
The authors thank D.\ Efremov, 
J.\ van den Brink,
D.V.\ Evtushinsky, V.B.\ 
 Zabolotnyy, S.\ Borisenko,  
and G.\ Prando for useful discussions as well as M.\ 
Deutschmann, S.\ M\"{u}ller-Litvanyi, 
R.\ M\"{u}ller, J.\ Werner, S.\ Pichl, K.\ Leger, 
and S.\ Gass for technical support.
This was supported by the DFG through SPP 1458
and Grants No. GR3330/2 and BE1749/13. SW acknowledges funding by DFG
in project WU 595/3-1.  S.J.\ 
 acknowledges financial support from
the Foundation for Fundamental Research on Matter (FOM, The Netherlands).
\end{acknowledgments}

\section*{Appendix A}
In adopting Eq.\ (\ref{product}) the ``true" electron-boson interaction is somewhat 
underestimated as compared with a more natural description \cite{Iwasawa2013}
\begin{equation}
\gamma_{\rm el}= \left(   
1+ \lambda_{\rm e-b}+\lambda_{\rm e-e} 
\right) \gamma_{\rm b} \ ,
\label{true}
\end{equation}
if $\lambda_{\rm e-e} >0$. 
Using the electron-electron (e-e) self-energy $\Sigma_{\rm e-e}(\omega)$, 
the true e-e coupling constant is given by
\begin{equation}
\lambda_{\rm e-e}= -\frac{\partial {\rm Re }\Sigma_{\rm e-e}}
{\partial \omega} \Big{|}_{\omega =0}. 
\label{lambdaelel}
\end{equation}
The advantage of adopting Eq.\ (\ref{product})
is the possibility to use standard Eliashberg-theory to extract 
$\lambda^{\rm eff}_{\rm b, tot}$ from the analysis of low-temperature
thermodynamical properties such as specific heat, penetration depth etc. 
If instead Eq.\ (\ref{true}) is used, phenomenological
model assumptions for the corresponding e-e 
self-energy $\Sigma_{\rm e-e}$ have to be adopted.\cite{Iwasawa2013} 
Iwasawa {\it et al.} (Ref. \onlinecite{Iwasawa2012}) proposed a simple expression
\begin{equation}
\Sigma_{\rm e-e}(\omega ) =\frac{g \omega }{\left(\omega^2 +i\gamma ^2\right)^2 } \quad ,
\end{equation}
where $g=0.5\beta\gamma^2$ with $\gamma \approx U_d$ (the screened on-site
Coulomb interaction) and $\beta$ as an 
empirical factor. Within this model one obtains from Eq.\ (\ref{lambdaelel})
\begin{equation}
\lambda_{\rm e-e}=0.5\beta U_d \quad .
\end{equation}
From the analysis of ARPES data for the 4$d$ oxide Sr$_2$RuO$_4$ these 
authors arrived
at $\lambda_{\rm e-e}= 1.6$ to 1.9 adopting $U_d=1.2$ to 1.5 eV and 
$\beta \approx 2.53$~eV$^{-1}$ for the latter value. In case of Fe pnictides
a slightly larger on-site Coulomb interaction is expected and we adopt 
$U_d \approx 2$~eV. Then with the same or a slightly enhanced $\beta $-value
of 3~eV$^{-1}$ one arrives at slightly larger e-e and e-b coupling constants 
$\lambda_{e-e}=2.53 \  {\rm to } \ 3 \ $ and 
$\lambda_{\rm e-b} \approx  1.5$ to 1, respectively, using the empirical
total mass enhancement of the order of five
obtained above in qualitative agreement also with
the standard Eliashberg-theory based analysis given in the main 
text.

\end{document}